\documentclass[12pt,twoside]{article}
\usepackage{fleqn,espcrc1,epsf}

\usepackage{graphicx}
\usepackage[figuresright]{rotating}

\newcommand{\AmS}{{\protect\the\textfont2
  A\kern-.1667em\lower.5ex\hbox{M}\kern-.125emS}}
\newcommand{\beq}{\begin{equation}}
\newcommand{\eeq}{\end{equation}}
\newcommand{\bea}{\begin{eqnarray}}
\newcommand{\eea}{\end{eqnarray}}

\newcommand{\sing}{$^1\!S_0$ }

\title{Describing Nuclear Matter with Effective Field Theories}

\author{James V. Steele
        and 
        R.J. Furnstahl\address{Department of Physics,
	The Ohio State University, Columbus, OH 43210 USA}%
        \thanks{Supported in part by NSF grants PHY-9511923 and PHY-9800964.}}
\begin{document}

\maketitle

\begin{abstract}
An accurate description of nuclear matter starting from free-space
nuclear forces has been an elusive goal.
The complexity of the system makes approximations inevitable,
so the challenge is to find a consistent truncation scheme with controlled
errors. 
The virtues of an effective field theory approach to this problem
are discussed.
\end{abstract}

\smallskip
\noindent\rule{2.5in}{.5pt}
\medskip

Nuclear forces have been studied in depth over the past fifty years,
leading to excellent phenomenological descriptions.
A recent resurgence of interest in this field has been fueled by
the promise of systematic results from an effective
field theory (EFT) analysis~\cite{eft}. 
EFT techniques
adapted to many-body systems may give new insight into the properties
of nuclei and their connection to QCD.

The EFT lagrangian consists of long-range interactions constrained by
chiral symmetry and the most general short-range interactions
consistent with QCD symmetries.
The coefficients of these short-range terms may eventually be derived
from QCD, 
but at present must be fit by matching calculated and experimental
observables in a momentum expansion.
This effective lagrangian
can then be used to systematically 
predict other observables, including inelastic processes. 

The predictability of an EFT relies on an organizational scheme,
called power counting, to determine the importance of any
contribution to a calculation. 
This can be illustrated by using the
familiar one-boson-exchange representation of the NN
force as a model of the true underlying physics.
One-pion exchange (OPE) is taken to be the long-range physics, as
its contribution to the EFT is dictated by chiral symmetry.
Exchanges of the heavier mesons will be considered short-range
physics, which must be included generically in the EFT. 
For example, for center-of-mass momenta below the sigma mass, $p<m_\sigma$, 
the heavy-meson propagators are 
well approximated by momentum-dependent contact interactions,
\beq
\frac1{p^2-m_\sigma^2} = C_0 + C_2 p^2 + C_4 p^4 + \ldots \ ,
\label{prop}
\eeq
with $C_2=C_0/m_\sigma^2$ and so on.
Each term has an associated power of
$p/m_\sigma$, which carries over to observables.
A consistent truncation based on counting these powers leads to observables
with well-defined errors \cite{eft}. 
Here $m_\sigma$ plays the role of the EFT breakdown scale $\Lambda$.
At momenta of order $\Lambda$,
the short-distance structure is resolved,
all higher-order corrections become comparable in magnitude,
and consequently the EFT fails.

In the early years of research into nuclear forces, Bethe showed that
low-energy scattering between two nucleons
could be expressed in a form now known as the effective range expansion.
For example, in the \sing channel,
\beq
p\cot\delta = - \frac1{a_s} + \frac12 r_e p^2 + \ldots \ .
\label{ere}
\eeq
A key observation is that the coefficients
$a_s$, $r_e$, $\ldots$,
 of this momentum expansion
are independent of the details of the potential used to fit 
 low-energy data.
The scale associated with the parameters is the point at which
Eq.~(\ref{ere}) breaks down, around $p\sim 1/r_e\sim m_\pi$.
This expansion
is suggestive of an EFT with a breakdown scale around 
$\Lambda \sim m_\pi$.
So we expect to reproduce Eq.~(\ref{ere}) if we treat {\it all\/}
interactions (including OPE) as short-ranged.

\begin{figure}
\vspace*{-.3in}
\begin{center}
\hbox{
\hspace*{2cm}
\epsfxsize=4in
\epsffile{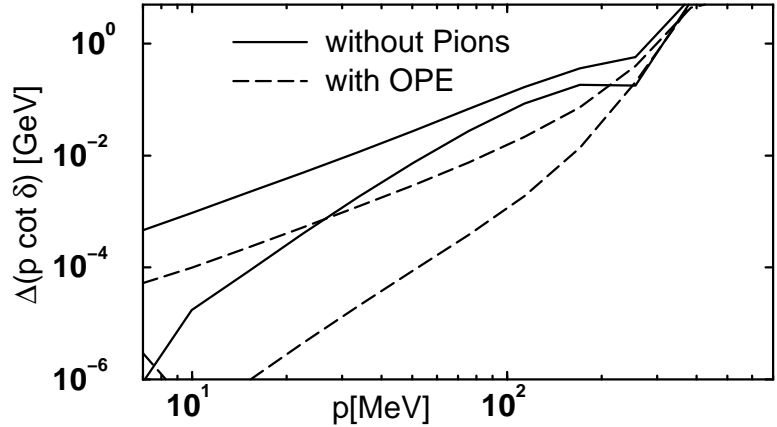}
}
\end{center}
\vspace{-.6in}
\caption{\label{fig:err} The error in $p\cot\delta$ for
\sing $np$ scattering using
cutoff regularization without pions (solid) and with one-pion exchange
(dashed), 
both shown for one and two constants~\protect\cite{steelef}.}
\vspace{-.2in}
\end{figure}

Due to the existence of nuclear bound states (such as the deuteron)
near threshold, an EFT analysis for multiple nucleons
requires a nonperturbative treatment.
One formulation is based on the 
Lippmann-Schwinger equation for the $T$-matrix \cite{eft}, 
$\hat T = \hat V + \hat V \hat G_0 \hat T$ .
The most general effective potential with no long-range physics
takes the form
\beq
\hat V_{\rm EFT} = C_0 + C_2 \hat p^2 + {\cal O}(\hat p^4/\Lambda^4) \ .
\label{veft}
\eeq
If the EFT phase shifts are calculated and compared with actual data,
we find that the constants indeed form a hierarchy analogous to
Eq.~(\ref{prop}), but with the relevant scale being set by the pion
mass $m_\pi$:
\beq
C_0 \sim \frac{4\pi}{M m_\pi} \ , \qquad\qquad
C_2 p^2 \sim \frac{4\pi}{M m_\pi} \frac{p^2}{m_\pi^2} \ ,
\qquad
\ldots \ .
\eeq
Enforcing this power counting order-by-order in $p^2$ reproduces the
effective range expansion Eq.~(\ref{ere}) and
generates the solid lines in the error plot of
Fig.~\ref{fig:err}. 
The error is dominated by the first omitted term and so behaves
like a power of $p^2$. 
As more contact interactions are included in the
effective potential Eq.~(\ref{veft}), the slope of the
error curves gets steeper.  
The point at which the linear parts of the curves 
converge~\cite{steelef} indicates the
breakdown scale $\Lambda$, which is indeed of order $m_\pi$
(actually $m_\pi/2$).

To push beyond this scale, pions 
must be explicitly included in the effective lagrangian as long-range
physics.
There are alternative power counting schemes proposed in the  
literature~\cite{eft}.
In Fig.~\ref{fig:err}, results are shown for a scheme in which
one-pion exchange enters at leading order and pion effects
are treated nonperturbatively.
The range of validity of the EFT is extended to about
$\Lambda=300$~MeV in that case, as seen by the dashed lines.

A complete calculation at next-to-leading order requires 
irreducible two-pion exchange.
Preliminary results suggest that with a careful removal of long-distance
contamination from coefficients by using a
modified effective range
expansion~\cite{steelef},
the
EFT range of validity may extend as high as
$\Lambda \sim m_\rho$~\cite{steelef2}.
Even then, the accuracy of the EFT results for NN scattering
is not yet competitive with that of the Bonn potential~\cite{steelef2}.
Nevertheless, the error plots manifest a systematic expansion
that allows realistic error estimates for scattering and other
observables, which are not available in conventional approaches.

How does this expansion and breakdown scale carry over to nuclear matter?
The complications of the many-body problem can obscure the connection,
but we can use a perturbative matching calculation to understand the
general correspondence.
To proceed,
model data can be generated from an exactly solvable potential with weak
coupling $\lambda$.
The EFT contact interactions is then fit by this data up
to a given order in $\lambda$; for example,
\beq
\hat T_{\rm EFT} = \hat V_{\rm EFT} +\hat V_{\rm EFT}\; 
\hat G_0 \; \hat V_{\rm EFT} + {\cal
O}(\lambda^3) \ . 
\label{pert}
\eeq
The regularization of divergences in the second term of
Eq.~(\ref{pert})
necessitates
a renormalization of the first term to ensure a cutoff independent
match.
Solving for the exact in-medium $T$-matrix $\hat\Gamma$ 
perturbatively and comparing with the EFT result 
using the effective potential
in Eq.~(\ref{veft}) verifies that the match
remains regulator independent to the same order as in free space, 
\beq
\hat\Gamma = \hat\Gamma_{\rm EFT} + {\cal O}(\lambda^3) + {\cal
O}(k_F^4/\Lambda^4)  \ .
\eeq
The truncation error from the EFT expansion in nuclear matter 
is a power of the Fermi momentum over the
free-space breakdown scale, $k_F/\Lambda$~\cite{lutz}.   
This is very encouraging, since phenomenologically successful
mean-field descriptions have a related expansion with
$\Lambda\simeq600$~MeV~\cite{neg}.
Thus if the free-space breakdown scale is indeed as large as 
$\Lambda \sim m_\rho$,
a useful EFT expansion of nuclear matter is likely. 

A nonperturbative EFT calculation of nuclear matter introduces many
complications compared to the free-space EFT. 
The analogue of scattering between two nucleons involves particles
propagating off the energy-shell ($E\ne p^2/M$) and the
in-medium $T$-matrix not only depends on the relative momentum, 
but also the total momentum.
Also, enforcing the Pauli exclusion principle greatly
complicates the momentum 
integrals.
However,
a sys-tematic EFT error analysis makes the
calculation worth the effort,  
and the EFT approach can also
provide new perspectives on old issues and practices.
Some examples:
i) Nuclear saturation with velocity-dependent potentials instead of
a hard core;
ii) Renormalization scheme independence of observables, such as the
binding energy, {\it requires\/} the sum of ladder diagrams, 
providing a new justification for this approximation;
iii) A self-consistent ladder calculation requires 
explicit three-body forces.  This last feature
might account for discrepancies in nuclear matter between
phase-equivalent two-body potentials \cite{neg}.

\begin{figure}[t]
\begin{center}
\epsfxsize=6in
\epsffile[124 621 558 701]{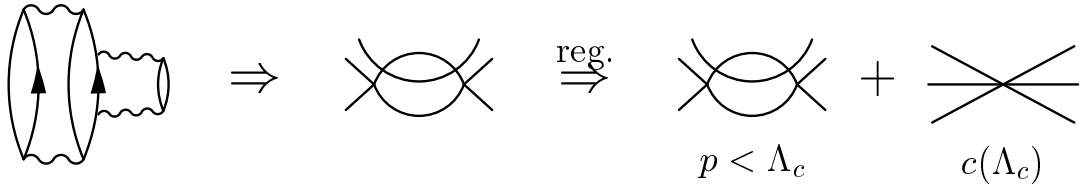}
\vspace{-.6in}
\end{center}
\caption{\label{fig:three}Regularization of 
a fourth-order contribution to the energy featuring 
only two-body interactions requires 
three-body contact interactions in the EFT.}
\vspace{-.2in}
\end{figure}

To see the necessity of three-body
forces, consider three-to-three scattering with only two-body 
interactions,
as depicted by the second Feynman diagram in Fig.~\ref{fig:three}.
This
could arise from the first diagram,
which is a contribution
to the energy of the nuclear matter
ground state,
by expanding the short-range interactions in contact terms and
opening the hole lines. 
Since the EFT is only accurate at low momenta, the
divergent loop integrals should be cut off at some scale $\Lambda_c$.
The intermediate states above $\Lambda_c$ 
are highly virtual, and so by the uncertainty principle are well
represented by  a regularized series of three-body 
contact interactions. 
These interactions are needed to 
guarantee regularization independence of the final result {\it and\/}
to account for contributions from the suppressed 
physical degrees of freedom.

Higher-body interactions are cut off in a similar
fashion, 
implying an infinite number of constants and many-body forces
would be needed to describe nuclear matter.
The Pauli exclusion principle helps the situation by limiting
the number of contact
interactions with no derivatives to four nucleons or fewer,
and the EFT power counting implies that other
higher-body interactions between nucleons are suppressed.  
However, the nature of the power counting is still under 
investigation, and it is 
possible that three- and four-body contact
terms are needed even at leading order~\cite{paulo} in a nuclear matter
calculation.

The inevitability of many-body forces in the EFT analysis
highlights the fact that the potential is not an observable.
There is no such thing as a ``best'' two-body potential
and differences in two-body off-shell behavior in nuclear matter
may be compensated by many-body counterterms~\cite{neg}.
In the case of three-nucleon scattering, 
accounting for the three-body contact interaction leads to a compelling
explanation of the Phillips line~\cite{paulo}.
An interesting possibility is that an extension of this analysis to
nuclear matter could explain the Coester line~\cite{machleidt}.

While the application of effective field theory to nuclear matter
is in its infancy,
systematic EFT calculations 
could help settle long-standing issues, justify traditional
expansions, or even offer new alternatives.
In the long run,
EFT techniques lay a foundation for ultimately connecting
the physics of nuclei to QCD.

\end{document}